\documentclass[a4paper,USenglish,cleveref]{article}

\usepackage{xcolor}
\definecolor{blue-vlight}  {rgb}{0.871, 0.937, 0.988}
\definecolor{red-vlight}   {rgb}{0.984, 0.871, 0.878}
\definecolor{myblue}         {rgb}{0.337, 0.592, 0.773}
\definecolor{myred}          {rgb}{0.776, 0.357, 0.396}

\usepackage{amsmath}
\usepackage{amssymb}
\usepackage{subcaption}
\usepackage[
colorlinks=true,
linkcolor=myred,
urlcolor=myblue,
citecolor=myblue,
linkbordercolor=blue-vlight,
urlbordercolor=blue-vlight,
pdftex
]{hyperref}
\usepackage{booktabs}

\title{Zipping Segment Trees\thanks{This work was partially funded by the Helmholtz Program ``Enery System Design'', Topic 2 ``Digitalization and System Technology''.}}

\author{\Large Lukas Barth\thanks{Karlsruhe Institute of Technology, Germany, \{firstname\}.\{lastname\}@kit.edu}
  \quad Dorothea Wagner\footnotemark[2]}

\newcommand{\thesisonly}[1]{}
\newcommand{\paperonly}[1]{#1}
\newcommand{\termdef}[1]{\emph{#1}}

\newcommand{\Om}{\ensuremath{\Omega}}

\usepackage{circledsteps}
\usepackage{todonotes}

\newcommand{\ie}{i.\,e.\xspace}

\newcommand{\R}{\ensuremath{\mathbb{R}}}

\newcommand{\Oh}{\ensuremath{\mathcal{O}}}

\newcommand{\Psrdm}[1]{\ifthenelse{\equal{#1}{*}}%
  {\textsc{SRDM}\xspace}%
  {\textsc{Scheduling with Release Times and Deadlines on a Minimum Number of Machines}\xspace#1}%
}
\newcommand{\Ptcpsp}[1]{\ifthenelse{\equal{#1}{*}}%
	{\textsc{TCPSP}\xspace}%
	{\textsc{Time-Constrained Project Scheduling Problem}\xspace#1}%
}
\newcommand{\Prcpsp}[1]{\ifthenelse{\equal{#1}{*}}%
	{\textsc{RCPSP}\xspace}%
	{\textsc{Resource-Constrained Project Scheduling Problem}\xspace#1}%
}
\newcommand{\Pracp}[1]{\ifthenelse{\equal{#1}{*}}%
	{\textsc{RACP}\xspace}%
	{\textsc{Resource Acquirement Cost Problem}\xspace#1}%
}
\newcommand{\Prip}[1]{\ifthenelse{\equal{#1}{*}}%
	{\textsc{RIP}\xspace}%
	{\textsc{Resource Investment Problem}\xspace#1}%
}

\newcommand{\Prlp}[1]{\ifthenelse{\equal{#1}{*}}%
	{\textsc{RIP}\xspace}%
	{\textsc{Resource Leveling Problem}\xspace#1}%
}

\newcommand{\Pfpsp}[1]{\ifthenelse{\equal{#1}{*}}%
	{\textsc{FPSP}\xspace}%
	{\textsc{Flexibilization Project Scheduling Problem}\xspace#1}%
}
\newcommand{\PfpspPs}[1]{\ifthenelse{\equal{#1}{*}}%
	{\textsc{FPSP-PS}\xspace}%
	{\textsc{Flexibilization Project Scheduling Problem with Peak Shaving}\xspace#1}%
}
\newcommand{\PfpspPsg}[1]{\ifthenelse{\equal{#1}{*}}%
	{\textsc{FPSP-PSG}\xspace}%
	{\textsc{Flexibilization Project Scheduling Problem with Peak Shaving and Generation}\xspace#1}%
}
\newcommand{\PfpspOm}[1]{\ifthenelse{\equal{#1}{*}}%
	{\textsc{FPSP-OM}\xspace}%
	{\textsc{Flexibilization Project Scheduling Problem with Overshoot Minimization}\xspace#1}%
}
\newcommand{\Psracp}[1]{\ifthenelse{\equal{#1}{*}}%
	{\textsc{S-RACP}\xspace}%
	{\textsc{Single-Resource Acquirement Cost Problem}\xspace#1}%
}
\newcommand{\Psromp}[1]{\ifthenelse{\equal{#1}{*}}%
	{\textsc{S-ROMP}\xspace}%
	{\textsc{Single-Resource Overshoot Minimization Problem}\xspace#1}%
}

\newcommand{\rush}[1][noopt]{\ifthenelse{\equal{#1}{noopt}}{RUSH\xspace}{RUSH$\langle #1 \rangle$\xspace}}
\newcommand{\frush}[1][noopt]{\ifthenelse{\equal{#1}{noopt}}{F-RUSH\xspace}{F-RUSH$\langle #1 \rangle$\xspace}}
\newcommand{\swag}[1][noopt]{\ifthenelse{\equal{#1}{noopt}}%
	{\textsc{SWAG}\xspace}%
	{\textsc{Scheduling With Augmented Graphs}\xspace}%
}
\newcommand{\grasp}[1][]{\ifthenelse{\equal{#1}{*}}{\textsc{Greedy
Randomized Adaptive Search Procedure}\xspace}{\textsc{GRASP}\xspace}}

\usepackage[ruled, vlined, linesnumbered, noend]{algorithm2e}
\usetikzlibrary{fit}

\definecolor{thesisblue-light}   {rgb}{0.490, 0.710, 0.867}
\colorlet{primarycolor-light}     {thesisblue-light}

\newcommand*{\tikzmk}[1]{\tikz[remember picture,overlay,] \node (#1) {};\ignorespaces}
\newcommand{\boxit}[1]{\tikz[remember picture,overlay]{\node[yshift=4pt,xshift=4pt,fill=#1,opacity=.25,fit={(A)(B)}] {};}\ignorespaces}

\begin{document}

\maketitle

\begin{abstract}
  Stabbing queries in sets of intervals are usually answered using segment trees. A dynamic variant
  of segment trees has been presented by van Kreveld and Overmars~\cite{kreveld1993union}, which
  uses red-black trees to do rebalancing operations. This paper presents \emph{zipping segment
    trees} --- dynamic segment trees based on zip trees, which were recently introduced by Tarjan et
  al.~\cite{tarjan2019zip}. To facilitate zipping segment trees, we show how to uphold certain
  segment tree properties during the operations of a zip tree. We present an in-depth experimental
  evaluation and comparison of dynamic segment trees based on red-black trees, weight-balanced trees
  and several variants of the novel zipping segment trees. Our results indicate that zipping segment
  trees perform better than rotation-based alternatives.
\end{abstract}

\section{Introduction}

A common task in computational geometry, but also many other fields of application, is the storage
and efficient retrieval of segments (or more abstractly: intervals). \thesisonly{In this thesis, we
  require such a data structure for the heuristics presented in chapters \ref{chap:rush} and
  \ref{chap:swag}, where we use it to localize the demand peak in a given schedule.} The question of
which data structure to use is usually guided by the nature of the retrieval operations, and whether
the data structure must by dynamic, i.e., support updates. \thesisonly{For scheduling algorithms, the concrete task is to determine all jobs that run at a
  specified time $t$. This amounts to a classic \termdef{stabbing query} in an interval
  storing data structure, }\paperonly{One very common retrieval operation is that of a
  \termdef{stabbing query}, } which can be formulated as follows: Given a set of
intervals on $\R$ and a query point $x \in \R$, report all intervals that contain $x$.

For the static case, a \termdef{segment tree} is the data structure of choice for this task. It
supports stabbing queries in $\Oh(\log n)$ time (with $n$ being the number of intervals). Segment
trees were originally introduced by Bentley~\cite{bentley1977algorithms}. While the segment
tree is a static data structure, i.e., is built once and would have to be rebuilt from scratch to
change the contained intervals, van Kreveld and Overmars present a dynamic version
\cite{kreveld1993union}, called \termdef{dynamic segment tree} (DST).

Dynamic segment trees are applied in many fields. Solving problems from computational geometry
certainly is the most frequent application, for example for route planning based on
geometry~\cite{edelkamp2005geometric} or labeling rotating maps~\cite{gemsa2016evaluation}. However,
DSTs are also useful in other fields, for example internet
routing~\cite{chang2007dynamic}\paperonly{ or scheduling algorithms~\cite{barth2019shaving}}.

\thesisonly{To apply the dynamic segment tree to scheduling problems, we treat a job in a schedule
  as an interval between its starting and finishing point, and associate the intervals with a
  weight, i.e., the (in the simplest case single-resource) demand of the corresponding job. We store
  these intervals in a dynamic segment tree that we extend with an additional annotation at every
  node. This annotation allows us to determine the peak demand as well as the time interval during
  which the peak demand occurs in $\Oh(1)$.}

In this \thesisonly{chapter}\paperonly{paper}, we present an adaption of dynamic segment trees,
so-called \termdef{zipping segment trees}. Our main contribution is replacing the usual
red-black-tree base of dynamic segment trees with zip trees, a novel form of balancing binary search
trees introduced recently by Tarjan et al.~\cite{tarjan2019zip}. On a conceptual level, basing
dynamic segment trees on zip trees yields an elegant and simple variant of dynamic segment
trees. Only few additions to the zip tree's rebalancing methods are necessary. On a practical
level, we can show that zipping segment trees outperform dynamic segment trees based on red-black trees in our experimental setting.

\section{Preliminaries}%
\label{dsts:sec:prelim}

A concept we need for zip trees are the two \termdef{spines} of a (sub-) tree. We also talk
about the spines of a node, by which we mean the spines of the tree rooted in the respective
node. The \emph{left spine} of a subtree is the path from the tree's root to the previous (compared
to the root, in tree order) node. Note that if the root (call it $v$) is not the overall
smallest node, the left spine exits the root left, and then always follows the right child, i.e., it
looks like $(v, L(v), R(L(v)), R(R(L(v))), \dots)$. Conversely, the \emph{right spine} is the path
from the root node to the next node compared to the root node. Note that this definition differs
from the definition of a spine by Tarjan et al.~\cite{tarjan2019zip}.

\subsection{Union-Copy Data Structure}%
\label{dsts:sec:unioncopy}

Dynamic segment trees in general carry annotations of sets of intervals at their vertices or
edges. These set annotations must be stored and updated somehow. To achieve the run times in
\cite{kreveld1993union}, van Kreveld and Overmars introduce the \termdef{union-copy} data structure
to manage such sets. Sketching this data structure would be out of scope for this
\thesisonly{chapter}\paperonly{paper}. It is constructed by intricately nesting two different types
of union-find data structures: a textbook union-find data structure using union-by-rank and path
compression (see for example Seidel and Sharir~\cite{seidel2005top}) and the $UF(i)$ data structure
by La Poutr{\'e}~\cite{lapoutre1989new}.

For this \thesisonly{chapter}\paperonly{paper}, we just assume this union-copy data structure to
manage sets of items. It offers the following operations\footnote{The data structure presented by
  Kreveld and Overmars provides more operations, but the ones mentioned here are sufficient for this \thesisonly{chapter}\paperonly{paper}.}:
\begin{description}
  \item[createSet()] Creates a new empty set in $\Oh(1)$.
  \item[deleteSet()] Deletes a set in $\Oh(1 + k \cdot F_N(n))$, where $k$ is the number of elements in the set, and $F(n)$ is the time the \emph{find} operation takes in one of the chosen union-find structures. 
  \item[copySet(A)] Creates a new set that is a copy of $A$ in $\Oh(1)$.
  \item[unionSets(A,B)] Creates a new set that contains all items that are in $A$ or $B$, in $\Oh(1)$.
  \item[createItem(X)] Creates a new set containing only the (new) item $X$ in $\Oh(1)$.
  \item[deleteItem(X)] Deletes $X$ from \emph{all} sets in $\Oh(1 + k)$, where $k$ is the number of sets $X$ is in.
\end{description}

\subsection{Dynamic Segment Trees}%
\label{dsts:sec:dsts}

This section recapitulates the workings of dynamic segment trees as presented by van Kreveld and
Overmars~\cite{kreveld1993union} and outlines some extensions. Before we describe the dynamic
segment tree, we briefly describe a classic static segment tree and the \termdef{segment tree
  property}. For a more thorough description, see de Berg et
al.~\cite[10.2]{deberg2008computational}. Segment trees store a set $\mathcal{I}$ of $n$
intervals. Let $x_1, x_2, \dots x_{2n}$ be the ordered sequence of interval end points in
$\mathcal{I}$. For the sake of clarity and ease of presentation, we assume that all interval borders
are distinct, i.e., $x_i > x_{i+1}$. We also assume all intervals to be closed. See
Section~\ref{sec:equal-keys} in the appendix for the straightforward way of dealing with equal interval borders as
well as arbitrary combinations of open and closed interval borders.

In the first step, we forget whether an $x_i$ is a start or an end of an interval. The intervals
$$(-\infty, x_1), [x_1, x_1], (x_1, x_2), [x_2, x_2], \dots (x_{2n - 1}, x_{2n}), [x_{2n}, x_{2n}], (x_{2n}, \infty)$$

are called the \termdef{elementary intervals} of $\mathcal{I}$. To create a segment tree, we create
a leaf node for every elementary interval. On top of these leaves, we create a binary tree. The
exact method of creating the binary tree is not important, but it should adhere to some balancing
guarantee to provide asymptotically logarithmic depths of all leaves.

\begin{figure}
  \centering
  \includegraphics{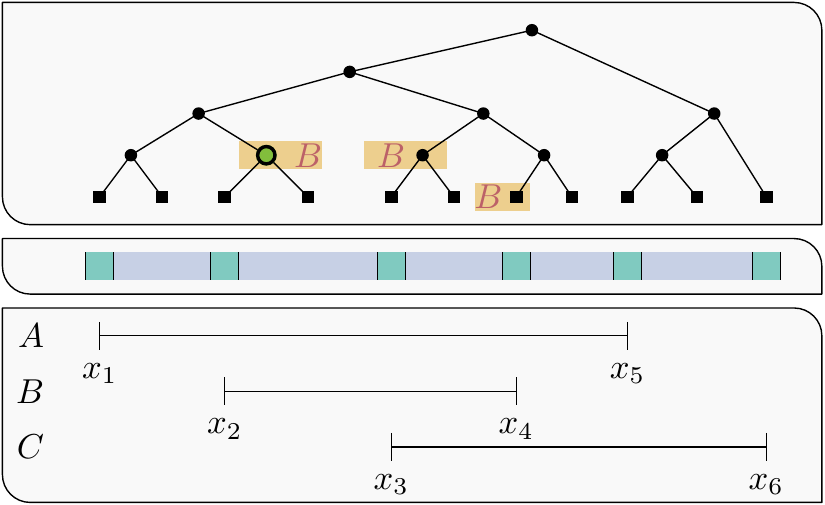}
  \caption{A segment tree (top) for three intervals (bottom). The middle shows the elementary
    intervals. Note that the green intervals do actually contain just one point and are only drawn
    fat so that they can be seen. The nodes marked with $B$ are the nodes that carry the annotation
    for interval $B$.}%
  \label{dsts:fig:segtree}
\end{figure}

Such a segment tree is outlined in Figure~\ref{dsts:fig:segtree}. The lower box indicates the three
stored intervals and their end points $x_1, \dots x_6$. The middle box contains a visualization of
the elementary intervals, where the green intervals are the $[x_i, x_i]$ intervals (note that while
of course they should have no area, we have drawn them ``fat'' to make them visible) while the blue
intervals are the $(x_i, x_{i+1})$ intervals. The top box contains the resulting segment tree, with
the square nodes being the leaf nodes corresponding to the elementary intervals, and the circular
nodes being the inner nodes.

We associate each inner node $v$ with the union of all the intervals corresponding to the leaves in
the subtree below $v$. In Figure~\ref{dsts:fig:segtree}, that means that the larger green inner node
is associated with the intervals $[x_2, x_3)$, \ie, the union of $[x_2, x_2]$ and $(x_2, x_3)$,
which are the two leaves beneath it. Recall that a segment tree should support fast stabbing queries,
i.e., for any query point $q$, should report which intervals contain $q$. To this end, we annotate
the nodes of the tree with sets of intervals. For any interval $I$, we annotate $I$ at every node
$v$ such that the associated interval of $v$ is completely contained in $I$, but the associated
interval of $v$'s parent is not. In Figure~\ref{dsts:fig:segtree}, the annotations for $B$ are
shown. For example, consider the larger green node. Again, its associated interval is $[x_2, x_3)$,
which is completely contained in $B = [x_2, x_4]$. However, its parent is associated with
$[x_1, x_3)$, which is not contained in $B$. Thus, the large green node is annotated with $B$.

A segment tree constructed in such a way is semi-dynamic. Segments cannot be removed, and new
segments can be inserted only if their end points are already end points of intervals in $I$. To
provide a fully dynamic data structure with the same properties, van Kreveld and Overmars present
the dynamic segment tree \cite{kreveld1993union}. It relaxes the property that intervals are always
annotated on the topmost nodes the associated intervals of which are still completely contained in
the respective interval. Instead, they propose the \termdef{weak segment tree property}: For any
point $q$ and any interval $I$ that contains $q$, the search path of $q$ in the segment tree
contains exactly one node that is annotated with $I$. For any $q$ and any interval $J$ that does not
contain $q$, no node on the search path of $q$ is annotated with $J$. Thus, collecting all
annotations along the search path of $q$ yields the desired result, all intervals that contain
$q$. It is easy to see that this property is true for segment trees: For any interval $I$ that
contains $q$, some node on the search path for $q$ must be the first node the associated interval of
which does not fully contain $I$. This node contains an annotation for $I$.

Dynamic segment trees also remove the distinction between leaf nodes and inner nodes. In a dynamic
segment tree, every node represents an interval border. To insert a new interval, we insert two
nodes representing its borders into the tree, adding annotations as necessary. To delete an
interval, we remove its associated nodes. If the dynamic segment tree is based on a classic
red-black tree, both operations require rotations to rebalance. Performing such a rotation without
adapting the annotations would cause the weak segment tree property to be violated. Also, the nodes
removed when deleting an interval might have carried annotations, which also potentially violates
the weak segment tree property.

We must thus fix the weak segment tree property during rotations. We must also make sure that any
deleted node does not carry any annotations, and we must specify how we add annotations when
inserting new intervals.





\section{Zipping Segment Trees}%
\label{dsts:sec:zipping-dst}

In Section~\ref{dsts:sec:dsts} we have described a variant of the dynamic segment trees introduced
by van Kreveld and Overmars~\cite{kreveld1993union}. These are built on top of a balancing binary
search tree, for which van Kreveld and Overmars suggested using red-black trees. The presented
technique is able to uphold the weak segment tree property during the red-black tree's operations:
node rotations, node swaps, leaf deletion and deletion of vertices of degree one. These are
comparatively many operations that must be adapted to dynamic segment trees. Also, each each
operation incurs a run time cost for the repairing of the weak segment tree property.

\begin{figure}
  \centering

  \begin{subfigure}[t]{0.25\textwidth}
    \includegraphics{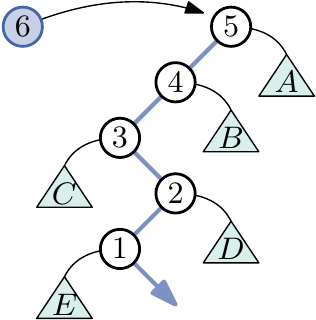}
    \caption{Before unzipping}%
    \label{dsts:fig:unzipping:before}
  \end{subfigure}
  \quad
  \begin{subfigure}[t]{0.25\textwidth}
    \includegraphics{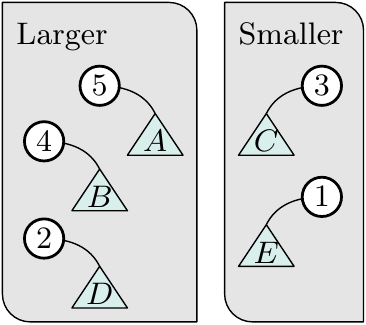}
    \caption{Unzipped parts}%
    \label{dsts:fig:unzipping:parts}
  \end{subfigure}
  \quad\quad
  \begin{subfigure}[t]{0.35\textwidth}
    \includegraphics{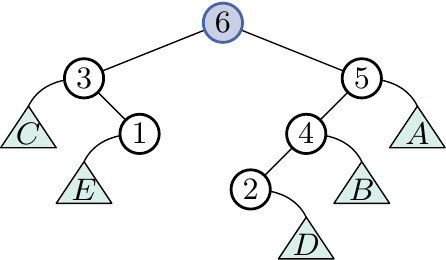}
    \caption{After reassembly}%
    \label{dsts:fig:unzipping:after}
  \end{subfigure}

  \caption{Illustration of the process of unzipping a path in zip trees. Nodes' names are
    simultaneously their ranks. Node keys are not shown.}%
  \label{dsts:fig:unzipping}
\end{figure}

Thus it stands to reason to look at different underlying trees which either reduce the number of
necessary balancing operations. One such data structure are \termdef{zip trees} introduced
by Tarjan et al.~\cite{tarjan2019zip}. Instead of inserting nodes at the bottom of the tree and then
rotating the tree as necessary to achieve balance, these trees determine the height of the node to
be inserted before inserting it in a randomized fashion by drawing a \termdef{rank}. The zip tree
then forms a \termdef{treap}, a combination of a search tree and a heap: While the key of $L(v)$
(resp. $R(v)$) must always be smaller or equal (resp.\ larger) to the key of $v$, the ranks of both
$L(v)$ and $R(v)$ must also be smaller or equal to the rank of $v$. Thus, any search path always
sees nodes' ranks in a monotonically decreasing sequence. The ranks are chosen randomly in such a
way that we expect the result to be a balanced tree. In a balanced binary tree, half of the nodes
will be leaves. Thus, we assign rank $0$ with probability $1/2$. A fourth of the nodes in a balanced
binary tree are in the second-to-bottom layer, thus we assign rank $1$ with probability $1/4$. In
general, we assign rank $k$ with probability $(1/2)^{k + 1}$, i.e., the ranks follow a geometric
distribution with mean $1$. With this, Tarjan et al.\ show that the expected length of search paths
is in $\Oh(\log n)$, thus the tree is expected to be balanced.

Zip trees do not insert nodes at the bottom or swap nodes down into a leaf before deletion. If nodes
are to be inserted into or removed from the middle of a tree, other operations than rotations are
necessary. For zip trees, these operations are \termdef{zipping} and \termdef{unzipping}. In the
remainder of this section, we examine these two operations of zip trees separately and explain how
to adapt them to preserve the weak segment tree property. For a more thorough description of the zip
tree procedures, we refer the reader to \cite{tarjan2019zip}.

\subsection{Insertion and Unzipping}

Figure~\ref{dsts:fig:unzipping} illustrates the unzipping operation that is used when inserting a
node. Note that we use the numbers $1$ through $6$ as nodes' names as well as their ranks in this
example. The triangles labeled $A$ through $E$ represent further subtrees. The node to be inserted
is \Circled{6}, the fat blue path is its search path (i.e., its key is smaller than the keys of
\Circled{5}, \Circled{4} and \Circled{2}, but larger than the keys of \Circled{3} and
\Circled{1}). Since \Circled{6} has the largest rank in this example, the new node needs to become
the new root. To this end, we unzip the search path, splitting it into the parts that are --- in
terms of nodes' keys --- larger than \Circled{6} and parts that are smaller than \Circled{6}. In
other words: We group the nodes on the search path by whether we exited them to the left (a larger
node) or to the right (a smaller node). Algorithm~\ref{dsts:alg:unzipping}, when ignoring the
highlighted parts, provides pseudocode for the unzipping operation.

\begin{algorithm}[t!]
  \SetKwInOut{KwInput}{Input}
  \KwInput{$new$: Node to be inserted}
  \KwInput{$v$: Node to be replaced by $new$}
  $cur \gets v$\;
  $oldParent \gets P(v)$\;
  $smaller \gets$ newList()\;
  $larger \gets$ newList()\;
  \tikzmk{A}$collected \gets \text{createSet()}$\tikzmk{B}\boxit{primarycolor-light}\;

  \tcc{Step 1: Remove edges along search path.}
  \While{$cur \neq \bot$}{\label{dsts:ln:loop-unzip}
    \eIf{$new < cur$} {
      $larger$.append($cur$)\;
      $next \gets L(cur)$\;
      \tikzmk{A}$S(\vec{R}(cur)) \gets \text{unionSets}(S(\vec{R}(cur)), collected)$\;\label{dsts:ln:add-other}
      $collected \gets \text{unionSets}(collected, S(\vec{L}(cur)))$\tikzmk{B}\boxit{primarycolor-light}\;\label{dsts:ln:collect}
      $next \gets L(cur)$\;
      $L(cur) \gets \bot$\;
      $cur \gets next$\;
    }{
      \tcc{Omitted, symmetric to the case $new < cur.$ Collect parts in $smaller$.}
    }
  }

  \tcc{Step 2: Insert $new$.}
  \eIf{$L(oldParent) = v$} {
    $L(oldParent) \gets new$\;
  } {
    $R(oldParent) \gets new$\;
  }

  \tcc{Step 3: Reassemble left spine from parts smaller than $new$}
  $parent \gets new$\;
  \For{$n \in smaller$} {\label{dsts:ln:loop-reassemble}
    \eIf{$parent = new$} {
      $L(parent) \gets n$\;
        \tikzmk{A}$\text{deleteSet}(S(\vec{L}(parent)));\> S(\vec{L}(parent)) \gets \text{createSet()}$\tikzmk{B}\boxit{primarycolor-light}\;\label{dsts:ln:empty-1}

    } {
      $R(parent) \gets n$\;
      \tikzmk{A}$\text{deleteSet}(S(\vec{R}(parent)));\> S(\vec{R}(parent)) \gets \text{createSet()}$\tikzmk{B}\boxit{primarycolor-light}\;\label{dsts:ln:empty-2}
    }
  }
  \tcc{Step 4: Reassemble right spine. This is symmetric to the left spine and thus omitted.}

  \caption{Unzipping routine. This inserts $new$ into the tree at the position currently occupied by $v$ by first disassembling the search path below $v$, and then reassembling the different parts as left and right spines below $new$. The highlighted parts are used to repair the dynamic segment tree's annotations. Note that in an efficient implementation, one would interleave all four steps.}%
  \label{dsts:alg:unzipping}
\end{algorithm}

We remove all edges on the search path (Step 1 in Algorithm~\ref{dsts:alg:unzipping}). The result is
depicted in the two gray boxes in Figure~\ref{dsts:fig:unzipping:parts}: several disconnected parts
that are either larger or smaller than the newly inserted node. Taking the new node \Circled{6} as
the new root, we now reassemble these parts below it. The smaller parts go into the left subtree of
\Circled{6}, stringed together as each others' right children (Step 3 in
Algorithm~\ref{dsts:alg:unzipping}). Note that all nodes in the ``smaller'' set must have an empty
right subtree, because that is where the original search path exited them --- just as nodes in the
``larger'' set have empty left subtrees. The larger parts go into the right subtree of \Circled{6},
stringed together as each others' left children. This concludes the unzipping operation, yielding
the result shown in Figure~\ref{dsts:fig:unzipping:after}. With careful implementation, the whole
operation can be performed during a single traversal of the search path.

To insert a segment into a dynamic segment tree, we need to do two things: First, we must correctly
update annotations whenever a segment is inserted. Second, we must ensure that the tree's
unzipping operation preserves the weak segment tree property.

We will not go into detail on how to achieve step one. In fact, we add new segments in basically the
same fashion as red-black-tree based DSTs do. We first insert the two nodes representing the
segment's start and end. Take the path between the two new nodes. The nodes on this path are the
nodes at which a static segment tree would carry the annotation of the new segment. Thus, annotating
these nodes (resp.\ the appropriate edges) repairs the weak segment tree property for the new
segment.

In the remainder of this section, we explain how to adapt the unzipping operations of zip trees to
repair the weak segment property. Let the annotation of an edge $e$ before unzipping be $S(e)$, and
let the annotation after unzipping be $S'(e)$. As an example how to fix the annotations after
unzipping, consider in Figure~\ref{dsts:fig:unzipping} a search path that descends into subtree $D$
before unzipping. It picks up the annotations on the unzipped path from \Circled{5} up to
\Circled{2}, i.e., $S\big(\vec{L}(5)\big)$, $S\big(\vec{L}(4)\big)$, $S\big(\vec{R}(3)\big)$, and on
the edge going into $D$, i.e., $S\big(\vec{R}(2)\big)$. After unzipping, it picks up the annotations
on all the new edges on the path from \Circled{6} to \Circled{2} plus $S'\big(\vec{R}(2)\big)$. We
set the annotations on all newly inserted edges to $\emptyset$ after unzipping. Thus, we need to add
the annotations before unzipping, i.e.,
$S\big(\vec{L}(5)\big) \cup S\big(\vec{L}(4)\big) \cup S\big(\vec{R}(3)\big)$, to the edge going
into $D$. We therefore set
$S'\big(\vec{R}(2)\big) = S\big(\vec{R}(2)\big) \cup S\big(\vec{L}(5)\big) \cup
S\big(\vec{L}(4)\big) \cup S\big(\vec{R}(3)\big)$ after unzipping.

In Algorithm~\ref{dsts:alg:unzipping}, the blue highlighted parts are responsible for repairing the
annotations. While descending the search path to be unzipped, we incrementally collect all
annotations we see on this search path (line~\ref{dsts:ln:collect}), and at every visited node add
the previously collected annotations to the other edge (line~\ref{dsts:ln:add-other}), \ie, the edge
that is not on the search path. By setting the annotations of all newly created edges to the empty
set (lines \ref{dsts:ln:empty-1} and \ref{dsts:ln:empty-2}), we make sure that after reassembly,
every search path descending into one of the subtrees attached to the reassembled parts picks up the
same annotations on the edge into that subtree as it would have picked up on the path before
disassembly.

\subsection{Deletion and Zipping}

Deleting segments again is a two-staged challenge: We need to remove the deleted segment from all
annotations, and must make sure that the \emph{zipping} operation employed for node deletion in zip
trees upholds the weak segment tree property. Removing a segment from all annotations is trivial
when using the union-copy data structure outlined in Section~\ref{dsts:sec:unioncopy}: The
\emph{deleteItem()} method does exactly this.

\begin{algorithm}[t!]
  \SetKwInOut{KwInput}{Input}
  \KwInput{$n$: Node to be removed}
  $l \gets L(n)$\tcp*{Current node descending $v$'s left spine}
  $r \gets R(n)$\tcp*{Current node descending $v$'s right spine}
  $p \gets P(n)$\tcp*{Bottom of the partially zipped path}
  $attachRight \gets R(P(n)) = v$\;
  \tikzmk{A}$collected_l = \text{copySet}(S(\vec{L}(n)))$\;
  $collected_r = \text{copySet}(S(\vec{R}(n)))$\tikzmk{B}\boxit{primarycolor-light}\;

  \While{$l \neq \bot \vee r \neq \bot$ \label{dsts:ln:zipping-loop}} {
    \eIf{$(l \neq \bot) \wedge ((r = \bot) \vee (\text{rank}(l) > \text{rank}(r)))$ \label{dsts:ln:zipping-selection}} {
      \eIf{attachRight} {
        $R(p) \gets l$\;
        \tikzmk{A}$\text{deleteSet}(S(\vec{R}(p)));\> S(\vec{R}(p)) \gets \text{createSet()}$\tikzmk{B}\boxit{primarycolor-light}\;
      } {
        $L(p) \gets l$\;
        \tikzmk{A}$\text{deleteSet}(S(\vec{L}(p)));\> S(\vec{L}(p)) \gets \text{createSet()}$\tikzmk{B}\boxit{primarycolor-light}\;
      }
      
      \tikzmk{A}$S(\vec{L}(l)) \gets \text{unionSets}(S(\vec{L}(l)), collected_l)$\;
      $collected_l \gets \text{unionSets}(collected_l, S(\vec{R}(l))$\tikzmk{B}\boxit{primarycolor-light}\;

      $p \gets l$\;
      $l \gets R(l)$\;
      $attachRight \gets \text{true}$\;
    } {\label{dsts:ln:zipping-right} 
      \eIf{attachRight} {
        $R(p) \gets r$\;
        \tikzmk{A}$\text{deleteSet}(S(\vec{R}(p)));\> S(\vec{R}(p)) \gets \text{createSet()}$\tikzmk{B}\boxit{primarycolor-light}\;
      } {
        $L(p) \gets r$\;
        \tikzmk{A}$\text{deleteSet}(S(\vec{L}(p)));\> S(\vec{L}(p)) \gets \text{createSet()}$\tikzmk{B}\boxit{primarycolor-light}\;
      }

      \tikzmk{A}$S(\vec{R}(r)) \gets \text{unionSets}(S(\vec{R}(r)), collected_r)$\;
      $collected_r \gets \text{unionSets}(collected_r, S(\vec{L}(r))$\tikzmk{B}\boxit{primarycolor-light}\;

      $p \gets r$\label{dsts:ln:zipping-rightparent}\;
      $r \gets L(r)$\;
      $attachRight \gets \text{false}$\label{dsts:ln:zipping-attachleft}\;
    }
  }
  \caption{Zipping routine. This removes $v$ from the tree, zipping the left and right spines of $v$. The highlighted parts are used to repair the dynamic segment tree's annotations.}%
  \label{dsts:alg:zipping}
\end{algorithm}

We now outline the \emph{zipping} procedure and how it can be amended to repair the weak segment tree property. Zipping two paths in the tree works in reverse to unzipping. Pseudocode is
given in Algorithm~\ref{dsts:alg:zipping}. Again, the pseudocode without the highlighted parts is
the pseudocode for plain zipping, independent of any dynamic segment tree. Assume that in the
situation Figure~\ref{dsts:fig:unzipping:after}, we want to remove \Circled{6}, thus we want to arrive
at the situation in Figure~\ref{dsts:fig:unzipping:before}. The zipping operation consist of
walking down the left spine (consisting of \Circled{3} and \Circled{1} in the example) and the right spine
(consisting of \Circled{5}, \Circled{4} and \Circled{2} in the example) simultaneously and zipping both into a single
path. This is done by the loop in line~\ref{dsts:ln:zipping-loop}. At every point during the walk,
we have a current node on both spines, call it the \emph{current left} node $l$ and the
\emph{current right} node $r$. Also, there is a \emph{current parent} $p$, which is the bottom of
the new zipped path being built. In the beginning, the current parent is the parent of the node
being removed.\footnote{If the root is being removed, pretend there is a pseudonode above the
  root.} In each step, we select the current node with the smaller rank, breaking ties arbitrarily
(line~\ref{dsts:ln:zipping-selection}). Without loss of generality, assume the current right node is
chosen (the branch starting in line~\ref{dsts:ln:zipping-right}). We attach the chosen node to the
bottom of the zipped path ($p$), and then $r$ itself becomes $p$. Also, we walk further down on the
right spine.

Note that the choice whether to attach left or right to the bottom of the zipped path (made via
$attachRight$ in Algorithm~\ref{dsts:alg:zipping}) is made in such a way that the position in which
we attach previously was part of one of the two spines being zipped. For example, if $p$ came from
the right spine, we attach left to it. However, $\vec{L}(p)$ comes from the right spine. This method
of attaching nodes always upholds the search tree property: When we make a node from the right spine
the new parent (line~\ref{dsts:ln:zipping-rightparent}), we know that the new $p$ is currently the
largest remaining nodes on the spines. We always attach left to this $p$
(line~\ref{dsts:ln:zipping-attachleft}). Since all other nodes on the spine are smaller than $p$,
this is valid. The same argument holds for the left spine.

We now explain how the edge annotations can be repaired so that the weak segment tree
property is upheld. Assume that for an edge $e$, $S(e)$ is the annotation of $e$
before zipping, and $S'(e)$ is the annotation of $e$ after zipping. Again, we argue via the subtrees
that search paths can descend into. A search path descending into a subtree on the right of a node
on the right spine, e.g., subtree $B$ attached to \Circled{4} in
Figure~\ref{dsts:fig:unzipping:after}, will before zipping pick up the annotation on
the right edge of the node being removed plus all annotations on the spine up to the respective
node, e.g., $S\big(\vec{R}(6)\big) \cup S\big(\vec{L}(5)\big)$, before descending into the
respective subtree ($B$ in the example). To preserve these picked up annotations, we again push them
down onto the edge that actually leads away from the spine into the respective subtree.

Formally, during zipping, we keep two sets of annotations, one per spine. In Algorithm~\ref{dsts:alg:zipping}, these are $collected_l$ and $collected_r$,
respectively. Let $n$ be the node to be removed. Initially, we set
$collected_l = S\big(\vec{L}(n)\big)$ and
$collected_r = S\big(\vec{R}(n)\big)$. Then, whenever we pick a node $c$ from the left
(resp.\ right) spine as new parent, we set
$S'\big(\vec{L}(c)\big) = S\big(\vec{L}(c)\big) \cup collected_l$
(resp.\
$S'\big(\vec{R}(c)\big) = S\big(\vec{R}(c)\big) \cup collected_r$). This
pushes down everything we have collected to the edge leading away from the spine at
$c$. Then, we set $collected_l = collected_l \cup S\big(\vec{R}(c)\big)$ and
$S'\big(\vec{R}(c)\big) = \emptyset$ (resp.\
$collected_r = collected_r \cup S\big(\vec{L}(c)\big)$ and
$S'\big(\vec{L}(c)\big) = \emptyset$). This concludes the techniques necessary to use zip trees as a basis for dynamic segment trees, yielding \emph{zipping segment trees}.

\subsection{Complexity}

Zip trees are randomized data structures, therefore all bounds on run times are expected bounds. In
\cite[Theorem 4]{tarjan2019zip}, Tarjan et al.\ claim that the expected number of pointers changed
during a zip or unzip is in $\Oh(1)$. However, they actually even show the stronger claim that the
number of nodes on the zipped (or unzipped) paths is in $\Oh(1)$. Observe that the loops in lines
\ref{dsts:ln:loop-unzip} and \ref{dsts:ln:loop-reassemble} of Algorithm~\ref{dsts:alg:unzipping} as
well as line \ref{dsts:ln:zipping-loop} of Algorithm~\ref{dsts:alg:zipping} are executed at most
once per node on the unzipping (resp.\ zipping) path. Inside each of the loops, a constant number of
calls are made to each of the \emph{copySet}, \emph{createSet}, \emph{deleteSet} and
\emph{unionSets} operations. Thus, the rebalancing operations incur expected constant effort plus
a constant number of calls to the union-copy data structure.

When inserting a new segment, we add it to the sets annotated at every vertex along the path between
the two nodes representing the segment borders. Since the depth of every node is expected
logarithmic in $n$, this incurs expected $\ln(n)$ calls to \emph{unionSets}. The deletion of a
segment from all annotations costs exactly one call to \emph{deleteItem}.

All operations but \emph{deleteSet} and \emph{deleteItem} are in $\Oh(1)$ if the union-copy data
structure is appropriately built. The analysis for the two deletion functions is more complicated
and involves amortization. The rough idea is that every non-deletion operation can
increase the size of the union-copy's representation only by a limited amount. On the
other hand, the two deletion operations each decrease the representation size proportionally to
their run time.

The red-black-tree-based DSTs by van Kreveld and Overmars~\cite{kreveld1993union} also need
$\Om(\ln n)$ calls to \emph{copySet} during the insertion operation, and at least a constant number
of calls during tree rebalancing and deletion. Therefore, for every operation on zipping segment
trees, the (expected) number of calls to the union-copy data structure's functions is no larger than
the number of calls in the red-black-tree-based implementation and we achieve the same (but only
expected) run time guarantees, which are $\Oh(\log n)$ for insertion, $\Oh(\log n \cdot a(i,n))$ for
deletion (with $a(i,n)$ being the row-inverse of the Ackermann function, for some constant
  $i$) and $\Oh(\log n + k)$ for stabbing queries, where $k$ is the number of reported segments.

\subsection{Generating Ranks}%
\label{dsts:sec:ranks}

Nodes' ranks play a major role in the rebalancing operations of zip trees. In
Section~\ref{dsts:sec:zipping-dst}, we already motivated why nodes' ranks should follow a geometric
distribution with mean $1$; it is the distribution of the node depths in a perfectly balanced
tree.

A practical implementation needs to somehow generate these values. The obvious implementation would
be to somehow generate a (pseudo-) random number and determine the position of the first $1$ in its
binary representation. The rank generated in this way is then stored at the respective node.

Storing the rank at the node can be avoided if the rank is generated in a reproducible
fashion. Tarjan et al.~\cite{tarjan2018zip_arxiv} already point out that one can ``compute it as a
pseudo-random function of the node (or of its key) each time it is needed.'' In fact, the idea
already appeared earlier in the work by Seidel and Aragon~\cite{seidel1996randomized} on
treaps. They suggest evaluating a degree 7 polynomial with randomly chosen coefficients at the
(numerical representation of) the node's key. However, the 8-wise independence of the random
variables generated by this technique is not sufficient to uphold the theoretical guarantees given
by Tarjan et al.~\cite{tarjan2018zip_arxiv}.

However, without any theoretical guarantees, a simpler method for reproducible ranks can be achieved
by employing simple hashing algorithms. Note that even if applying universal hashing, we do not get
a guarantee regarding the probability distribution for the values of individual bits of the hash
values. However, in practice, we expect it to yield results similar to true randomness. As a fast
hashing method, we suggest using the $2/m$-almost-universal multiply-shift method from
Dietzfelbinger et al.~\cite{dietzfelbinger1997reliable}. Since we are interested in generating an
entire machine word in which we then search for the first bit set to $1$, we can skip the ``shift''
part, and the whole process collapses into a simple multiplication.

\section{Experimental Evaluation of Dynamic Segment Trees Bases}%
\label{dsts:sec:evaluation}

In this section, we experimentally evaluate zipping segment trees as well as dynamic segment trees
based on two of the most prominent rotation-based balanced binary search trees: red-black trees and
weight-balanced trees. Weight-balanced trees require a parametrization of their rebalancing
operation. In \thesisonly{Chapter~\ref{chap:engineering_wbts}, we
  perform}\paperonly{\cite{barth2020engineering}, we perform} an in-depth engineering
of weight-balanced trees. For this analysis of dynamic segment trees, we pick only the two most
promising variants of weight-balanced trees: top-down weight-balanced trees with
$\langle \Delta, \Gamma \rangle = \langle 3, 2 \rangle$ and top-down weight-balanced trees with
$\langle \Delta, \Gamma \rangle = \langle 2, 3/2 \rangle$.

Note that since we are only interested in the performance effects of the trees underlying the DST,
and not in the performance of an implementation of the complex union-copy data structure, we have
implemented the simplified variant of DSTs outlined in Section~\ref{dsts:sec:numeric} in the
appendix, which alters the concept of segment trees to only report the aggregate value of weighted
segments at a stabbing query, instead of a list of the respective segments. Evaluating the
performance of the union-copy data structure is out of scope of this work.

\begin{figure}[p]
  \begin{subfigure}{\textwidth}
    \centering
    \includegraphics{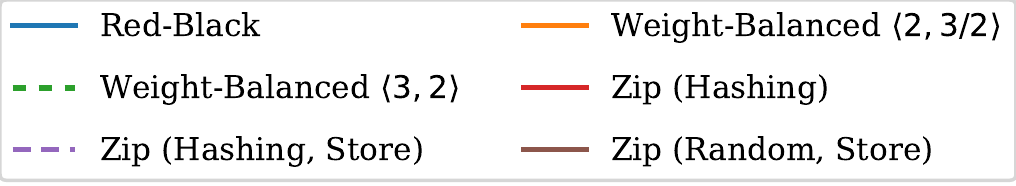}
  \end{subfigure}\\[.5em]
  \begin{subfigure}[t]{.48\textwidth}
    \includegraphics[scale=0.92]{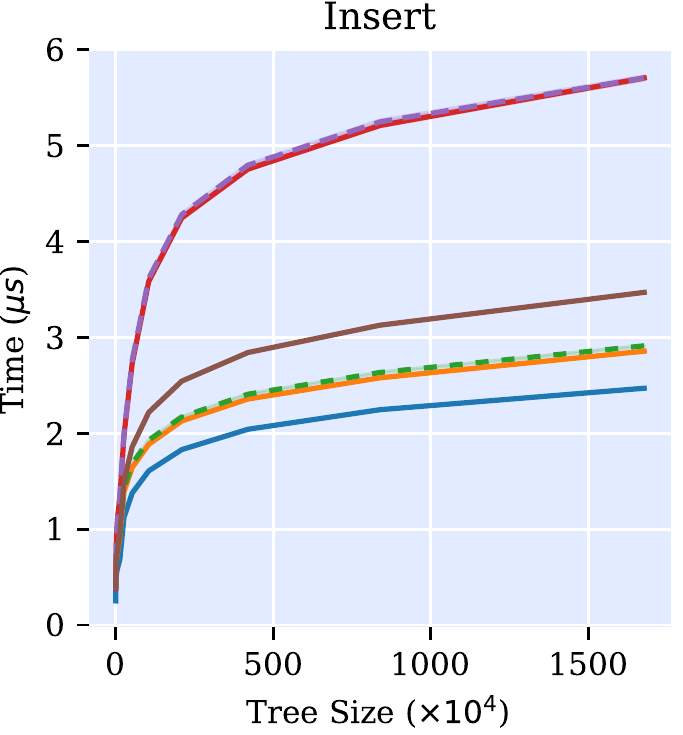}
    \caption{Timings for inserting a segment.}%
    \label{dsts:plot:ops:insert}
  \end{subfigure}
  \begin{subfigure}[t]{.48\textwidth}
    \includegraphics[scale=0.92]{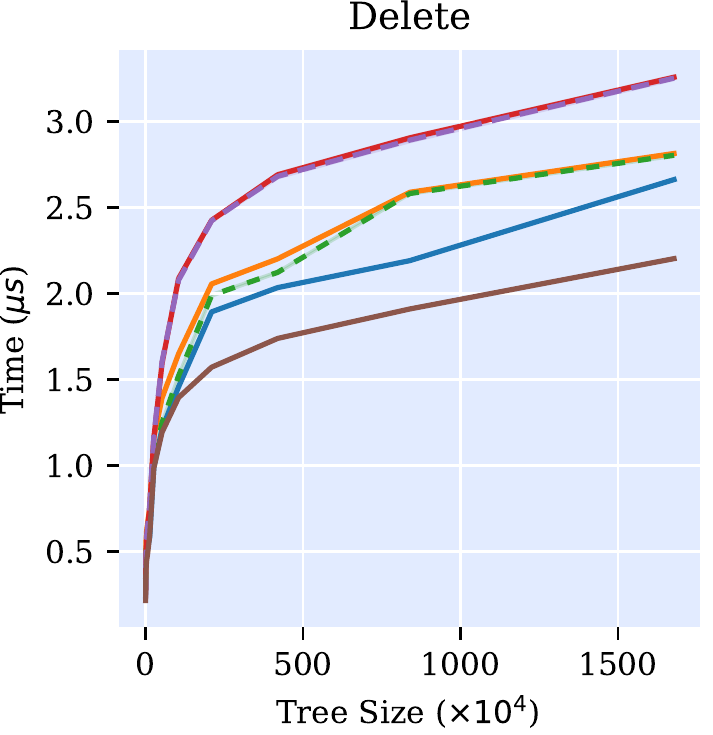}
    \caption{Timings for deleting a segment.}%
    \label{dsts:plot:ops:delete}
  \end{subfigure}

  \begin{subfigure}[t]{.48\textwidth}
    \includegraphics[scale=0.92]{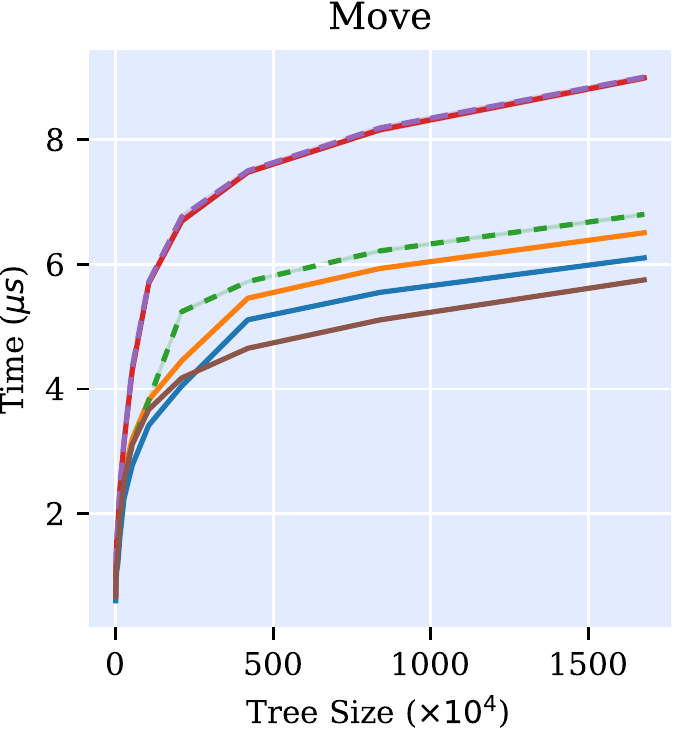}
    \caption{Timings for moving a segment.}%
    \label{dsts:plot:ops:move}
  \end{subfigure}
  \begin{subfigure}[t]{.48\textwidth}
    \includegraphics[scale=0.92]{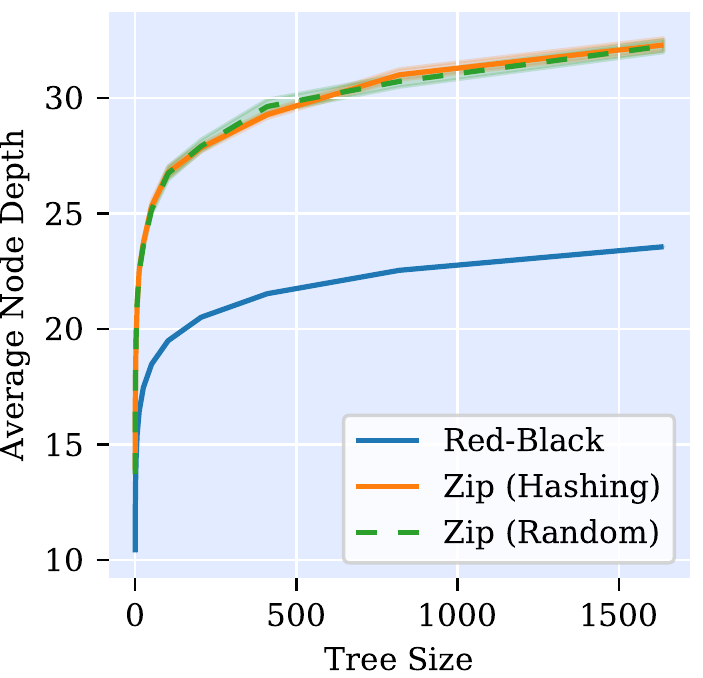}
    \caption{Average node depths.}%
    \label{dsts:plot:avgdepth}
  \end{subfigure}

    \caption{(a)---(c): Benchmark times for dynamic segment trees based on different balancing binary search trees. The $y$ axis indicates the measured time per operation, while the $x$ axis indicates the size of the tree that the operation is performed on. The lines indicate mean values. The standard deviation is all cases too small to be visible.\\
  (d): Average depths of the nodes in DSTs based on red-black trees and zip trees. The $x$ axis specifies the number of inserted segments. Shaded areas indicate the standard deviation.}%
  \label{dsts:plot:ops}
\end{figure}

For the zip trees, we choose a total of three variants, based on the choices explained in
Section~\ref{dsts:sec:ranks}: The first variant, denoted \emph{Hashing}, generates nodes' ranks by
applying the fast hashing scheme by Dietzfelbinger et al.~\cite{dietzfelbinger1997reliable} to the
nodes' memory addresses. In this variant, node ranks are not stored at the nodes but re-computed on
the fly when they are needed. The second variant, denoted \emph{Hashing, Store} also generates
nodes' ranks from the same hashing scheme, but stores ranks at the nodes. The last variant, denoted
\emph{Random, Store} generates nodes' ranks independent of the nodes and
stores the ranks at the nodes.

We first individually benchmark the two operations of inserting (resp.\ removing) a segment to
(resp.\ from) the dynamic segment tree. Our benchmark works by first creating a base dynamic segment
tree of a certain size, then inserting new segments (resp.\ removing segments) into that tree. The
number of new (resp.\ removed) segments is chosen to be the minimum of $10^5$ and $5\%$ of the base
tree size. Segment borders are chosen by drawing twice from a uniform distribution. All segments are
associated with a real-valued value, as explained in Section~\ref{dsts:sec:numeric}. We conduct our
experiments on a machine equipped with 128 GB of RAM and an Intel\textsuperscript{\textregistered}
Xeon\textsuperscript{\textregistered} E5-1630 CPU, which has 10 MB of level 3 cache. We compile
using GCC 8.1, at optimization level ``-O3 -ffast-math''. We do not run experiments concurrently. To
account for randomness effects, each experiment is repeated for ten different seed values, and
repeated five times for each seed value to account for measurement noise. All our code is published,
see Section~\ref{sec:code} in the appendix.

Figure~\ref{dsts:plot:ops:insert} displays the results for the insert operation. We see that the
red-black tree performs best for this operation, about a $30 \%$ faster ($\approx 2.5 \mu{}s$ per
operation at $1.5 \cdot 10^7$ nodes) than the fastest zip tree variant, which is the variant using
random rank selection ($\approx 3.5 \mu{}s$ per operation). The two weight-balanced trees lie
between the red-black tree and the randomness-based zip tree. Both hashing-based zip trees are
considerably slower.

For the deletion operation, shown in Figure~\ref{dsts:plot:ops:delete}, the randomness-based zip
tree is significantly faster than the best competitor, the red-black tree. Again, the
weight-balanced trees are slightly slower than the red-black tree, and the hashing-based zip trees
fare the worst.

Since (randomness-based) zip trees are the fastest choice for deletion and red-black trees are the
fastest for insertion, benchmarking the combination of both is obvious. Also, using an
\emph{dynamic} segment tree makes no sense if only the insertion operation is needed. Thus, we next
benchmark a \emph{move} operation, which consists of first removing a segment from the tree,
changing its borders, and re-inserting it. The results are shown in
Figure~\ref{dsts:plot:ops:move}. We see that the randomness-based zipping segment tree is the
best-performing dynamic segment tree for trees with at least $2.5 \cdot 10^6$ segments.


The obvious measurement to explore why different trees perform differently is the trees' balance,
i.e., the average depth of a node in the respective trees. We conduct this experiment as follows:
For each of the trees under study, we create trees of various sizes with randomly generated
segments. In a tree generated in this way, we only see the effects of the \emph{insert} operation,
and not the \emph{delete} operation. Thus, we continue by moving each segment once by removing it,
changing its interval borders and re-inserting it. This way, the effect of the \emph{delete}
operation on the tree balance is also accounted for. Since the weight-balanced trees were not
competitive previously, we perform this experiment only for the red-black and zip trees. We repeat the experiment with $30$ different seeds to account for randomness. The results can be found in
Figure~\ref{dsts:plot:avgdepth}. We can see that zipping segment trees, whether based on randomness
or hashing, are surprisingly considerably less balanced than red-black-based DSTs. Also, whether
ranks are generated from hashing or randomness does not impact balance.

Concluding the evaluation, we gain several insights. First, deletions in zipping segment trees are
so much faster than for red-black-based DSTs that they more than make up for the slower insertion,
and the fastest choice for moving segments are zipping segment trees with ranks generated
randomly. Second, we see that this speed does not come from a better balance, but in spite of a
worse balance. The speedup must therefore come from more efficient rebalancing operations. Third,
and most surprising, the question of how ranks are generated does not influence tree balance, but
has a significant impact on the performance of deletion and insertion. However, the hash function we
have chosen is very fast. Also, during deletion, no ranks should be (re-) generated for the variant
that stores the ranks at the nodes. Thus, the performance difference can not be explained by the
slowness of the hash function. Generating ranks with our chosen hash function must therefore
introduce some disadvantageous structure into the tree that does not impact the average node depth.

\section{Conclusion}

We have presented zipping segment trees --- a variation of dynamic segment trees, based on zip
trees. The technique to maintain the necessary annotations for dynamic segment trees is
comparatively simple, requiring only very little adaption  of zip trees' routines.  In our
 evaluation, we were able to show that zipping segment trees perform well in practice,
and outperform red-black-tree or weight-balanced-tree based DSTs with regards to
modifications. 

However, we were not yet able to discover exactly why generating ranks from a (very simple) hash
function does negatively impact performance. Exploring the adverse effects of this hash function and
possibly finding a different hash function that avoids these effects remains future work. Another
compelling future experiment would be to evaluate the performance when combined with the actual
 data structure by van Kreveld and Overmars.

All things considered, their relatively simple implementation and the superior performance when
modifying segments makes zipping segment trees a good alternative to classical dynamic segment trees
built upon rotation-based balancing binary trees.


\bibliography{references}

\clearpage
\appendix

\section{General Interval Borders}%
\label{sec:equal-keys}

In Section~\ref{dsts:sec:dsts}, we made two assumptions as to the nature of the intervals' borders:
we did assume a total ordering on the keys of the nodes, i.e., no segment border may appear in two
segments, and we assumed all intervals to be right-open. We now briefly show how to lift this
restriction.

The important aspect is that a query path must see the nodes representing interval borders on the
correct side. As an example, consider two intervals $I_1 = [a, b)$ and $I_2 = [b, c]$. A query for
the value $b$ should return $I_2$ but not $I_1$. Thus, the node representing ``$b)$'' must lie to
the left of the resulting search path, such that the query path does not end at a leaf between the
nodes representing ``$[a$'' and ``$b)$''. That way, the annotation for $I_1$ will not be picked
up. Conversely, the node representing ``$[b$'' must lie to the left of the query path, such that the
query path ends in a leaf between the nodes representing ``$[b$'' and ``$c]$''.

This dictates the ordering of nodes with the same numeric key $k$: First come the open upper borders,
then the closed lower borders, then the closed upper borders, and finally the open lower
borders. When querying for a value $k$, we descend left on nodes representing ``$k]$'' and ``$(k$'',
and descend right on nodes representing ``$[k$'' and ``$k)$''. That way, a search path for $k$ will
end at a leaf after all closed lower borders and open upper borders, but before all closed upper
borders and open lower borders. This yields the desired behavior.

\section{Numeric Annotations}%
\label{dsts:sec:numeric}
Many applications of segment-storing data structures deal with weighted segments, i.e., each segment
is associated with a number (or a vector of numbers). In such scenarios, one is often only
interested in determining the aggregated weight of the segments overlapping at a certain point
instead of the actual set of segments. 

This question can be answered without the need for a complicated union-copy data structure. In this
case, we annotate each edge with a real number resp.\ with a vector.  Instead of adding the actual
segments to the $S(\cdot)$ sets at edges, we just add the associated weight of the segment. The copy
operation is a simple duplication of a vector, a union is achieved by vector addition.

Deletion becomes a bit more complicated in this setting. Previously, we have exploited the
convenient operation of deleting an item from all sets offered by the union-copy data
structure. Now, say an interval associated with a weight vector $d \in \R^k$ is deleted
from the dynamic segment tree, and the segment is reperesented by the two nodes $a$ and $b$. If we
had just inserted the interval (and therefore $a$ and $b$), we would now add $d$ to the annotations
on a certain set of edges (see above for a description of the insertion process). When deleting an
interval, we annotate the same set of edges with $-1 \cdot d$. This exactly cancels out the
annotations made when the interval was inserted.

\section{Code Publication}%
\label{sec:code}
We publish our C++17 implementation of all evaluated tree variants, including all code to replicate
our benchmarks, at \url{https://github.com/tinloaf/ygg}. Note that this data structure library is
still work in progress and might change after the publication of this work. The exact version used
to produce the benchmarks shown in this paper can be accessed at
\url{https://github.com/tinloaf/ygg/releases/tag/version\_sea2020}.

The repository contains build instructions. The executables to generate our benchmark results can be
found in the \texttt{benchmark} subfolder. The \texttt{bench\_dst\_*} executables produce the
respective benchmark. The output format is that of Google Benchmark, see
\url{https://github.com/google/benchmark} for documentation.

The exact commands we have run to generate our data are:

\begin{center}
  \texttt{./bench\_DST\_Insert --seed\_start 42 --seed\_count 10 --benchmark\_repetitions=5 --benchmark\_out=<output\_dir>/result\_Insert.json --benchmark\_out\_format=json --doublings 14 --relative\_experiment\_size 0.05 --experiment\_size 100000}
  
  \texttt{./bench\_DST\_Delete --seed\_start 42 --seed\_count 10 --benchmark\_repetitions=5 --benchmark\_out=<output\_dir>/result\_Delete.json --benchmark\_out\_format=json --doublings 14 --relative\_experiment\_size 0.05 --experiment\_size 100000}
  
  \texttt{./bench\_DST\_Move --seed\_start 42 --seed\_count 10 --benchmark\_repetitions=5 --benchmark\_out=<output\_dir>/result\_Move.json --benchmark\_out\_format=json --doublings 14 --relative\_experiment\_size 0.05 --experiment\_size 100000}

\end{center}


\end{document}